\documentclass[final]{svjour3}
\usepackage{graphicx}
\usepackage{epsfig}
\usepackage{rotating}
\usepackage{amssymb}
\usepackage{mathptmx}
\usepackage[numbers]{natbib}
\makeatletter
\journalname{Journal of Low Temperature Physics}

\newcommand{\cecoin}    {CeCoIn$_5$}

\bibpunct{}{}{,}{s}{}{,}

\begin{document}

\title{Hyperfine fields and magnetic structure in the B phase of \cecoin}

\author{Nicholas J. Curro$^1$ \and Ben-Li Young$^2$ \and Ricardo R. Urbano$^3$ \and Matthias J. Graf$^4$}

\institute{
1: Department of Physics, University of California, Davis, CA 9516, USA; \email{curro@physics.ucdavis.edu}\\
2: Department of Electrophysics, National Chiao Tung University, Hsinchu 30010, Taiwan; \email{blyoung@mail.nctu.edu.tw}\\
3: National High Magnetic Field Laboratory, Florida State University, Tallahassee, Florida 32306-4005, USA; \email{urbano@magnet.fsu.edu}\\
4: Theoretical Division, Los Alamos National Laboratory, Los Alamos, New Mexico 87545, USA; \email{graf@lanl.gov}
}

\date{\today}

\maketitle

\begin{abstract}
We re-analyze Nuclear Magnetic Resonance (NMR) spectra observed at low temperatures and high magnetic fields
in the field-induced B-phase of \cecoin. The NMR spectra are consistent with incommensurate antiferromagnetic order of the Ce magnetic moments.
However, we find that the spectra of the In(2) sites depend critically on the direction of the ordered moments, the ordering wavevector and the symmetry of the hyperfine coupling to the Ce spins.
Assuming isotropic hyperfine coupling, the NMR spectra observed for $\mathbf{H}~||~[100]$ are consistent with magnetic order with wavevector
$\mathbf{Q}=\pi(\frac{1+\delta}{a},\frac{1}{a},\frac{1}{c})$
and Ce moments ordered antiferromagnetically along the [100] direction in real space.
If the hyperfine coupling has dipolar symmetry, then the NMR spectra require  Ce moments along the [001] direction.
The dipolar scenario is also consistent with recent neutron scattering measurements that find an ordered moment of 0.15$\mu_B$ along [001]
and
$\mathbf{Q_n}=\pi(\frac{1+\delta}{a},\frac{1+\delta}{a},\frac{1}{c})$
with incommensuration $\delta = 0.12$ for field $\mathbf{H}~||~[1\bar{1}0]$.
Using these parameters, we find that a hyperfine field with dipolar contribution is consistent with findings from both experiments.
We speculate that the B phase of \cecoin\ represents an intrinsic phase of modulated superconductivity and antiferromagnetism that can only emerge in a highly clean system.
\keywords{NMR \and superconductivity \and heavy fermion \and magnetism}
\PACS{P76.60.-k \and  75.30.Fv \and 74.10.+v}
\end{abstract}

\section{Introduction}
\label{intro}
The heavy-fermion superconductor \cecoin\ has attracted considerable attention since its discovery in 2001.\cite{ceirindiscovery}  Not only does this unconventional d-wave superconductor exhibit non-Fermi liquid behavior associated with proximity to a proposed quantum critical point,
but it is unique among the heavy-fermion superconductors in that it also exhibits a new thermodynamic phase
(B phase) that exists only within the superconducting phase near $H_{c2}$.\cite{sidorov,romanQCPCeCoIn5}
Initially this B phase was identified as the  elusive Fulde-Ferrell-Larkin-Ovchinnikov (FFLO) superconducting phase first predicted to exist
in Pauli-limited superconductors over 40 years ago.\cite{andrea,romanFFLO,radovanCeCoIn5FFLO,kakuyanagiCeCoIn5FFLO,kumagaiCeCoIn5FFLO}
In fact, recent NMR work by Young and coworkers\cite{CurroCeCoIn5FFLO}
identified the presence of incommensurate antiferromagnetic order in the B phase in contrast to the standard predictions for the FFLO
phase.\cite{Vorontsov2005,Vorontsov2006}
Signatures of magnetism were also seen in other NMR experiments.\cite{Mitrovic06, Koutroulakis08}
The NMR spectra of the In sites in the B phase do not reveal additional paramagnetic resonances that might
be associated with macroscopic phase separation, but rather are consistent with
superconducting order coexisting with a homogeneous modulation of antiferromagnetic order.\cite{CurroCeCoIn5FFLO}
Despite initial arguments to the contrary,\cite{MatsudaFFLOReview}
 recent neutron scattering results by Kenzelmann and coworkers now provide conclusive proof for long-range static incommensurate antiferromagnetic order.\cite{KenzelmannCeCoIn5Qphase}

\begin{figure*}
\begin{minipage}[t]{0.49\textwidth}
\centerline{  \includegraphics[width=\textwidth]{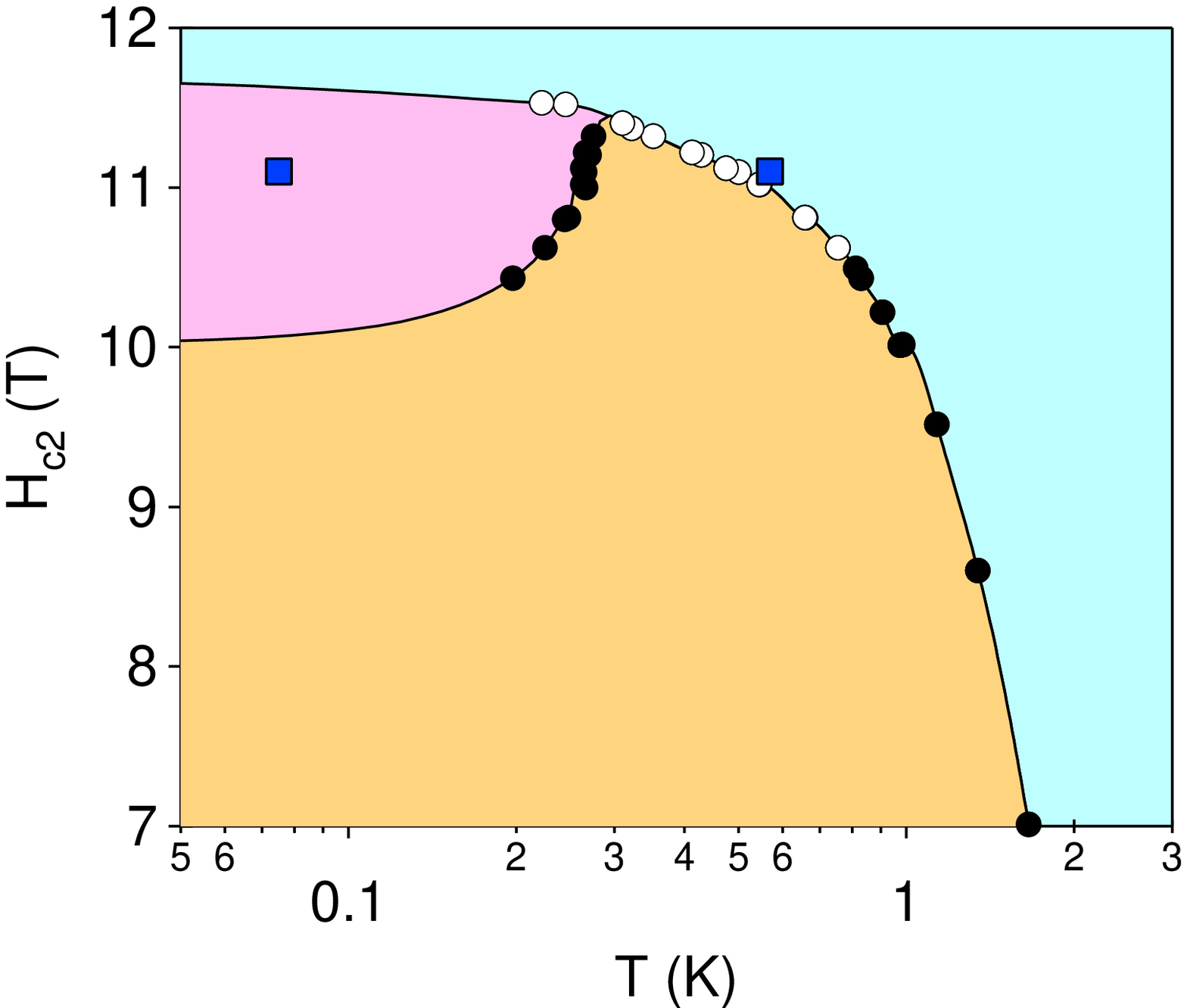} }
\caption{(Color online) The phase diagram of \cecoin\ in high field as determined by specific heat.\cite{romanFFLO}
Solid points represent second order phase transitions and open points are first order transitions.  The solid blue squares are the points at which the spectra in Fig. \ref{fig:spectra} were obtained.}
\label{fig:phasediagram}
\end{minipage}
\hfill
\begin{minipage}[t]{0.49\textwidth}
\centerline{  \includegraphics[width=0.7\textwidth]{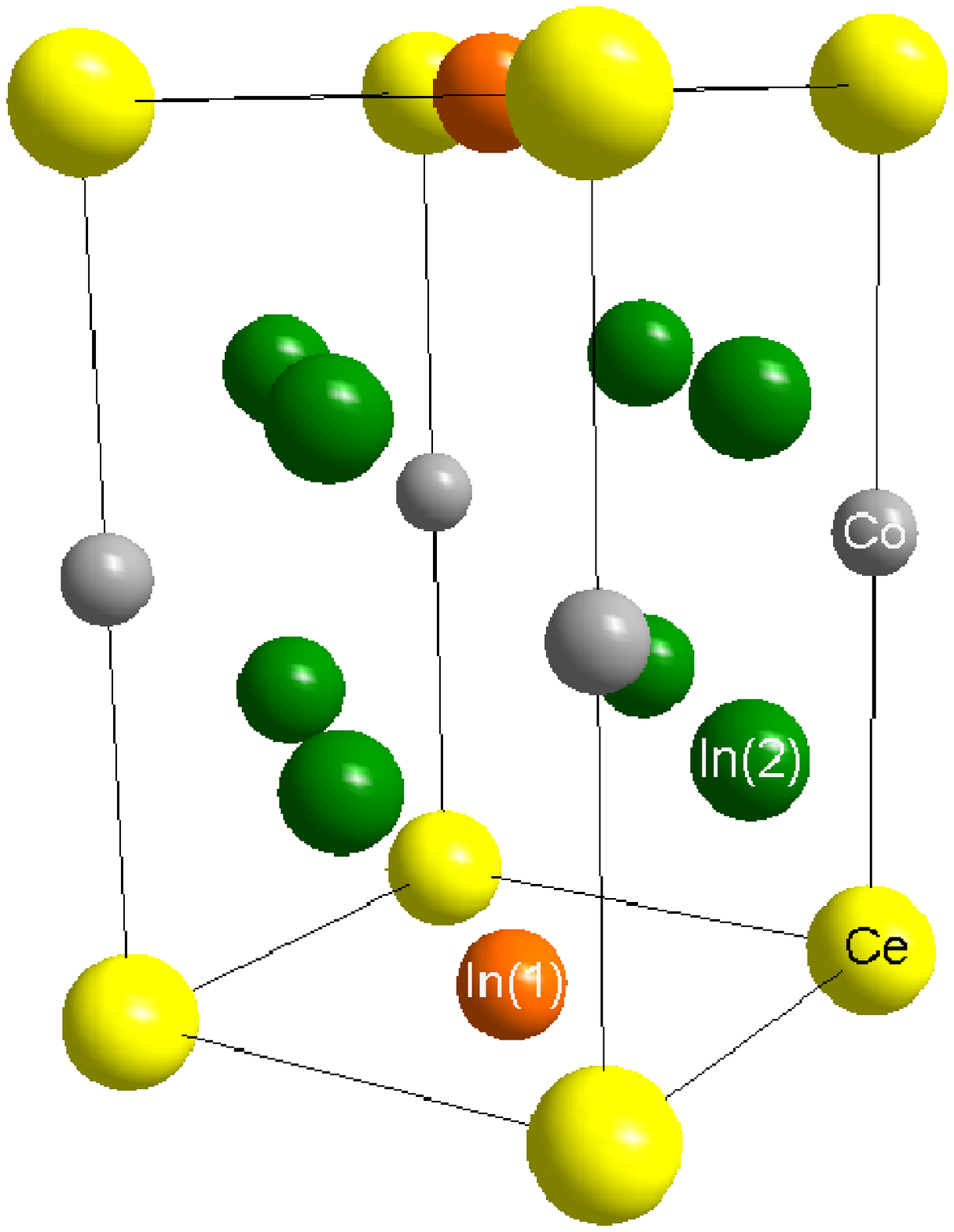} }
\caption{(Color online) The unit cell of \cecoin.  The Ce atoms (yellow) sit at the eight corners.
The In(1) atoms sit in the center of the top and bottom faces (orange).
The Co atoms are grey and the In(2) atoms are green.
For the field oriented in the $ab$ plane, there are two inequivalent In(2) atoms,
depending on whether the field is parallel (In(2a) or perpendicular (In(2b)) to the unit cell face.}
\label{fig:unitcell}
\end{minipage}
\end{figure*}

\begin{figure*}
\begin{center}
\includegraphics[width=0.9\textwidth]{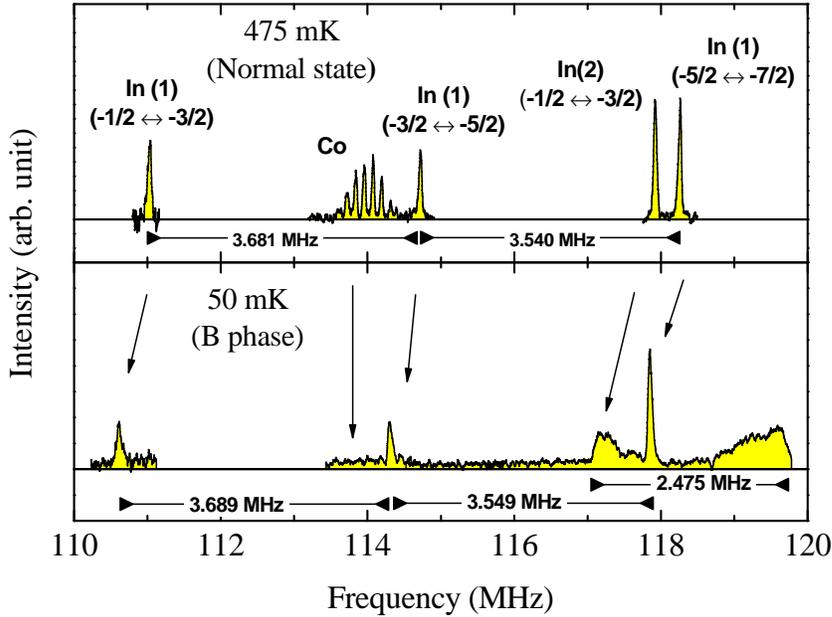}
\end{center}
\caption{(Color online) Fixed field
NMR spectra at 11.1 T in \cecoin\ showing how the In(1), In(2) and Co sites evolve from the normal state (top) to the B phase (bottom),
adapted from Young et al.\cite{CurroCeCoIn5FFLO} }
\label{fig:spectra}
\end{figure*}

The antiferromagnetism in \cecoin\ was first identified by Young et al.\
due to the presence of a broad spectrum observed at the In(2) sites in this material
(see Figs. \ref{fig:unitcell} and \ref{fig:spectra}).
The In(1) and Co sites, in contrast, showed no splitting.
Young et al.\  pointed out that these observations place constraints on the possible magnetic structure, but do not uniquely identify the structure.
They proposed a minimal model where the magnetic structure consists of ordered local Ce spins with moments $\mathbf{S}_0$ along the applied magnetic field direction (along [100]),
with an ordering wavevector of the form
$\mathbf{Q}=\pi(\frac{1+\delta}{a},\frac{1}{a},\frac{1}{c})$.
The structure of the NMR spectra revealed the incommensurate nature, but the value of the modulation $\delta$ remained undetermined since
the hyperfine field at the In(2) site depends on the product of the size of the ordered moment and the incommensuration.
In the neutron diffraction experiment, Kenzelmann et al.\ oriented the field along the [1$\bar{1}$0] direction,
and observed $\mathbf{Q_n} = \pi(\frac{1+\delta}{a},\frac{1+\delta}{a}, \frac{1}{c})$
with $\delta=0.12$ and moments along [001].  A crucial observation was that $\delta$ is \textit{independent of the applied field} in the B phase, in contrast to the predictions for the FFLO phase.
By proposing a Ginzburg-Landau model for the coupling of antiferromagnetism and superconductivity,
they showed that the superconducting order parameter in the B phase acquires a component with finite momentum
because of the strong coupling between the incommensurate magnetism and superconductivity.
The neutron data confirm the NMR observation that this exotic state disappears immediately above $H_{c2}$, where the system returns to a fully homogeneous normal phase, yet with strong deviations from conventional Fermi liquid
theory.\cite{romanQCPCeCoIn5}

A priori, the NMR and neutron scattering results suggest different magnetic structures.
In order to address this discrepancy, we investigate several possible magnetic structures allowed by the NMR results.
The neutron diffraction results suggest that the applied field, $\mathbf{H}$, the moments, $\mathbf{S}_0$, and the incommensuration wavevector $\mathbf{Q}_{i} = \frac{\pi}{a}(\delta, \delta, 0)$ are all mutually orthogonal.
In contrast, the proposed NMR scenario suggested $\mathbf{S}_0 ~||~\mathbf{H}~||~\mathbf{Q}_0$. As we show below,
this scenario is the most likely for \textit{isotropic} transferred hyperfine couplings between the Ce spins and the In(2) nuclei.
On the other hand, if the coupling tensors are \textit{anisotropic}, then other magnetic structures are possible as argued by Koutroulakis et al.\cite{Koutroulakis09, Mitrovic09}
If the hyperfine tensor has purely dipolar symmetry,
then we find that for $\mathbf{H}~||~[100]$ the most likely magnetic structures
satisfy $\mathbf{S}_0 ~||~ [001]$ and $\mathbf{Q}_i \perp \mathbf{S}_0$.  The spectra of the In(1) and In(2) differ slightly depending on the orientation of $\mathbf{Q}_i$ in the plane, but the data are most consistent with $\mathbf{Q}_i || [010]$ or $\mathbf{Q}_i || [110]$.  The anisotropic coupling scenario offers a picture that is both physically more reasonable and
consistent with neutron diffraction observations of
$\mathbf{Q}_i \perp \mathbf{S}_0$ and $\mathbf{H}_0 \perp \mathbf{S}_0$, though for fields $\mathbf{H}_0~||~[1\bar{1}0]$.

\section{Analysis}

\subsection{Spectra}

In order to explain the broad double-peak structure of the In(2) spectrum in Fig. \ref{fig:spectra}, there must be a distribution of local fields both parallel and antiparallel to the applied field $\mathbf{H}_0 || [100]$ with values ranging up to 1.3 kOe.\cite{CurroCeCoIn5FFLO}
The resonance frequency is given by $f = \gamma|\mathbf{H_0} + \mathbf{H}_{hf}| + f_Q$, where $\gamma$ is the gyromagnetic ratio and $f_Q$ is the contribution from the quadrupolar interaction at the nucleus.  Since the electric field gradient (EFG) at all three nuclear sites is unaffected by the onset of superconductivity or magnetism, the dramatic line broadening effects observed in the B phase can be attributed entirely to the onset of the static hyperfine field, $\mathbf{H}_{hf}$. $f_Q$ is a temperature independent constant that depends on the particular site and we will not address it further. Experimentally, we find no significant broadening in the B phase for the In(1) or the Co, but a broad, double-peak spectrum for the In(2a) site (previously referred to as In(2)$_{||}$, the In(2) site on the unit cell face that lies parallel to the field).  Independent measurements show no significant broadening at the In(2b) (previously referred to as In(2)$_{\perp}$).  These results put stringent constraints on any candidate magnetic structure.

\begin{figure*}
\begin{center}
\centerline{  \includegraphics[width=0.7\textwidth]{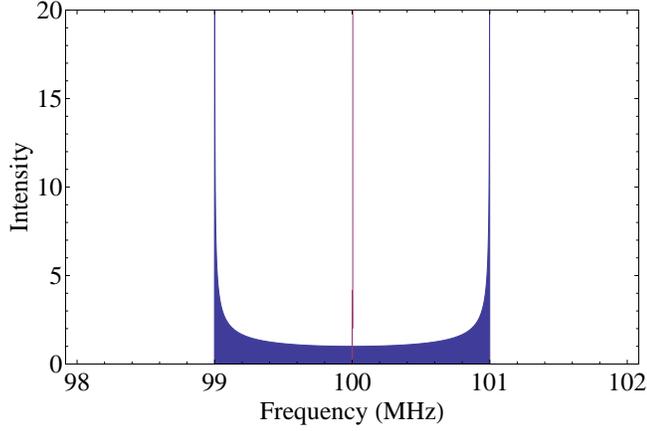} }
\end{center}
\caption{(Color online) The calculated spectra of $\mathcal{P}_{||}(f)$ (blue) and $\mathcal{P}_{\perp}(f)$ (red)
assuming $f_0$= 100 MHz and $\gamma h_{hf}^0$ = 1 MHz.
For the perpendicular case, the spectrum is only weakly affected by the hyperfine fields,
whereas for the parallel case it broadens dramatically resulting in a double-peak structure.}
\label{spectrumcalc}
\end{figure*}

The double-peak structure observed for the In(2a) arises because there is a distribution of local hyperfine fields that lie either parallel or antiparallel to the applied field.  If $\mathbf{H}_{hf}$ is parallel to $\mathbf{H}_0$ and is modulated along a direction $\hat{r}$, then $f(r) = f_0 + \gamma h_{hf}^0\cos(qr)$, where $h_{hf}^0$ is the magnitude of the modulation, $r$ is the distance along the modulation, and $q$ is the wavevector. In this case, the spectrum will then be given by
 \begin{equation}
 \mathcal{P}_{||}(f) \propto |df/dr|^{-1} =\frac{q^{-1}}{\sqrt{\gamma^2(h_{hf}^0)^2-(f-f_0)^2}}.
 \end{equation}
 On the other hand, if $\mathbf{H}_{hf}(r)~\perp  ~ \mathbf{H}_0$  then $f(r)=\sqrt{f_0^2+\gamma^2(h_{hf}^0)^2\cos^2(qr)}$ and the spectrum is given by
 \begin{equation}
 \mathcal{P}_{\perp}(f) \propto  \frac{q^{-1}f}{\sqrt{f^2-f_0^2}\sqrt{\gamma^2(h_{hf}^0)^2-f^2-f_0^2}}.
 \end{equation}
These spectra are shown in Fig. \ref{spectrumcalc}, and clearly show that for the parallel case,
there are two double peaks at frequencies both below and above $f_0$, whereas for the perpendicular case, there is only a single peak at higher frequency.
A key result of the re-analysis of the NMR spectra is, in agreement with our earlier analysis, that the In(2a) sites require a
hyperfine field parallel to the applied magnetic field,
$\mathbf{H}_{hf} ~||~ \mathbf{H}_0$, to account for the broadening and double-peak spectrum.
On the other side, the In(2b) sites show little or no broadening, consistent with a vanishing or perpendicular hyperfine field,
i.e., at the In(2b) sites either $\mathbf{H}_{hf}=0$ or $\mathbf{H}_{hf} \perp \mathbf{H}_{0}$.

\subsection{Hyperfine couplings}

The hyperfine interaction is given by the Hamiltonian
\begin{equation}
\mathcal{H}_{hf} = \mathbf{\hat{I}}\cdot \mathbb{A} \cdot\mathbf{S}(\mathbf{r}=\mathbf{0}) + \sum_i   \mathbf{\hat{I}}\cdot \mathbb{B}_i\cdot \mathbf{S}(\mathbf{r}_i),
\label{eq:hyptransferred}
\end{equation}
where the hyperfine coupling tensor $\mathbb{A}$ represents the on-site coupling to an electron spin,
$\mathbf{S}$, at the nuclear site $\mathbf{r}=0$,
$\mathbb{B}$ represents a transferred hyperfine coupling to an electron spin on a distant (ligand) site at $\mathbf{r}_i$.\cite{CurroKSA}
In \cecoin, these sites are the nearest neighbor Ce 4$f$ electrons, and the sum is over the nearest neighbors.  For static ordering of the Ce spins,
Eq.~(\ref{eq:hyptransferred}) can be re-written as: $\mathcal{H}_{hf} = \gamma\hbar \mathbf{\hat{I}}\cdot\mathbf{H}_{hf}$, where the magnitude and direction of the hyperfine field $\mathbf{H}_{hf}$ depend critically on the hyperfine tensors for the particular site and the magnetic structure.
The tensorial $\mathbb{A}$ term represents hyperfine coupling to the itinerant conduction electrons,
which we will ignore, since we are only concerned with static contributions to $\mathbf{H}_{hf}$ from the static local Ce ordering.

The transferred hyperfine tensor $\mathbb{B}$ is generally not diagonal in the crystal axis basis,\cite{Vachon08}
and may be written as the sum of isotropic and dipolar contributions.\cite{currohyperfine}
To lowest order, the tensor can be approximated by a scalar (isotropic) interaction, since the transferred hyperfine interaction is typically at least one order of magnitude greater than the direct dipolar interaction.
However, in the \cecoin\ compound, there is evidence that the hyperfine interaction is not purely isotropic, and therefore we must consider dipolar symmetries as well.
Indeed, the magnitude of the dipolar portion is found to be enhanced by the delocalized nature of the electrons in the solid.\cite{CPSbook}
Therefore, we write the hyperfine fields at the ligand sites as:
\begin{eqnarray}
  \mathbf{H}_{hf}(\mathbf{r}) &=& \sum_{i=1}^{4}\mathbb{B}_i\cdot\mathbf{S}(\mathbf{r}+\mathbf{r_i})/\gamma\hbar \mbox{~~~~~~~~~ at In(1) } \\
 \mathbf{H}_{hf}(\mathbf{r}) &=& \sum_{j=1}^{2}\mathbb{B}_i\cdot\mathbf{S}(\mathbf{r}+\mathbf{r_j})/\gamma\hbar \mbox{~~~~~~~~~ at Co } \\
  \mathbf{H}_{hf}(\mathbf{r}) &=& \sum_{k=1}^{2}\mathbb{B}_i\cdot\mathbf{S}(\mathbf{r}+\mathbf{r_k})/\gamma\hbar \mbox{~~~~~~~~~ at In(2a) }\\
  \mathbf{H}_{hf}(\mathbf{r}) &=& \sum_{l=1}^{2}\mathbb{B}_i\cdot\mathbf{S}(\mathbf{r}+\mathbf{r_l})/\gamma\hbar, \mbox{~~~~~~~~~ at In(2b) }
\end{eqnarray}
where $\mathbf{r}_i=( \pm \frac{a}{2},\pm \frac{a}{2},0 )$ for the In(1) nearest neighbor Ce sites,
$\mathbf{r}_j=( 0,0,\pm \frac{c}{2} )$ for the Co nearest neighbor Ce sites, and
$\mathbf{r}_l=( \pm \frac{a}{2},0,z_0 )$ for the In(2a) nearest neighbor Ce sites,
and $\mathbf{r}_k=( 0,\pm \frac{a}{2},z_0 )$ for the In(2b) nearest neighbor Ce sites. The couplings $\mathbb{B}_i = \mathbb{B}_{\rm iso} + \mathbb{B}_{\rm dip}$,
where:
\begin{equation}
\label{eqn:isotropic}
\mathbb{B}_{\rm iso} = B_{\rm iso}\left(
\begin{array}{ccc}
 1 & 0 & 0 \\
 0 & 1 & 0 \\
 0 & 0 & 1
\end{array}
\right)
\end{equation}
and
\begin{equation}
\label{eqn:dipolar}
    \mathbb{B}_{\rm dip} =
\frac{B_{\rm dip}}{r^2}\left(
\begin{array}{ccc}
 2 x^2-y^2-z^2 & 3 x y & 3 x z \\
 3 x y & -x^2+2 y^2-z^2 & 3 y z \\
 3 x z & 3 y z & -x^2-y^2+2 z^2
\end{array}
\right)
\end{equation}
are evaluated for each site.  Here $\mathbf{r} = (x,y,z)$ is the vector joining the particular site to the Ce atom in question.

\begin{table}[tbh]
  \caption{The hyperfine fields at the In(1) and In(2) sites for various magnetic structures.
Here we assumed that the applied magnetic field is $\mathbf{H}_0~||~[100]$,
the moment is 0.15 $\mu_B$ and its modulation is $\delta=0.12$.
The column at the far right indicates magnetic structures in agreement with the NMR spectra.}
  \label{fieldtable}
\begin{center}
\begin{tabular}{cccccccc}
 \hline\noalign{\smallskip}
Case  & $\mathbf{Q}_i$ & $\mathbf{S}_0$ & $\mathbb{B}$ & In(1) & In(2a) & In(2b) & Agreement? \\
  \hline\hline
  (1.1) & [100] & [100] & iso & $\mathbf{H}_{\rm hf} =0$ & $\mathbf{H}_{\rm hf} ~|| ~[100]$ & $\mathbf{H}_{\rm hf} =0$ & \textbf{yes (a)}\\
  (1.2)  & [100] & [100] & dip & $\mathbf{H}_{\rm hf}~ || ~[010]$ & $\mathbf{H}_{\rm hf} ~||~ [001]$ & $\mathbf{H}_{\rm hf} =0$ & no\\
  (1.3)  & [100] & [001] & iso & $\mathbf{H}_{\rm hf} =0$ & $\mathbf{H}_{\rm hf} =0$ & $\mathbf{H}_{\rm hf}~ ||~ [001]$ & no\\
  (1.4)  & [100] & [001] & dip & $\mathbf{H}_{\rm hf} =0$ & $\mathbf{H}_{\rm hf} ~||~[100]$ & $\mathbf{H}_{\rm hf} ~||~ [010]$ & \textbf{yes (b)}\\
  \hline
  (2.1) & [010] & [100] & iso & $\mathbf{H}_{\rm hf} =0$ & $\mathbf{H}_{\rm hf} =0$ & $\mathbf{H}_{\rm hf}~ || ~[100]$ & no\\
  (2.2) & [010] & [100] & dip & $\mathbf{H}_{\rm hf} ~|| ~[010]$ & $\mathbf{H}_{\rm hf} ~|| ~[001]$ & $\mathbf{H}_{\rm hf} \approx 0$ & no\\
  (2.3) & [010] & [001] & iso & $\mathbf{H}_{\rm hf} =0$ & $\mathbf{H}_{\rm hf} ~|| ~[001]$ & $\mathbf{H}_{\rm hf} =0$ & no\\
  (2.4) & [010] & [001] & dip & $\mathbf{H}_{\rm hf} =0$ & $\mathbf{H}_{\rm hf} ~||~ [100] $ & $\mathbf{H}_{\rm hf} ~||~ [010]$ & \textbf{yes (c)}\\
  \hline
  (3.1) & [110] & [100] & iso & $\mathbf{H}_{\rm hf} ~||~ [100]$ & $\mathbf{H}_{\rm hf}~ ||~ [100]$ & $\mathbf{H}_{\rm hf}~ ||~ [100]$ & no \\
  (3.2) & [110] & [100] & dip & $\mathbf{H}_{\rm hf} ~|| ~[001]$ & $\mathbf{H}_{\rm hf}~ ||~ [100]$ & $\mathbf{H}_{\rm hf}~ ||~ [010]$ & no\\
  (3.3) & [110] & [001] & iso & $\mathbf{H}_{\rm hf}~ ||~ [001]$ & $\mathbf{H}_{\rm hf} ~|| ~[001]$ & $\mathbf{H}_{\rm hf}~ || ~[001]$ & no\\
  (3.4) & [110] & [001] & dip & $\mathbf{H}_{\rm hf}~ ||~ [010]$ & $\mathbf{H}_{\rm hf}~ ||~ [100]$ & $\mathbf{H}_{\rm hf}~ ||~ [010]$  & \textbf{yes (d)}\\
  \hline
  (4.1) & [1$\bar{1}$0] & [100] & iso & $\mathbf{H}_{\rm hf} ~||~ [100]$ & $\mathbf{H}_{\rm hf}~ ||~ [100]$ & $\mathbf{H}_{\rm hf}~ ||~ [100]$ & no \\
  (4.2) & [1$\bar{1}$0] & [100] & dip & $\mathbf{H}_{\rm hf} ~|| ~[001]$ & $\mathbf{H}_{\rm hf}~ ||~ [100]$ & $\mathbf{H}_{\rm hf}~ ||~ [010]$ & no\\
  (4.3) & [1$\bar{1}$0] & [001] & iso & $\mathbf{H}_{\rm hf}~ ||~ [001]$ & $\mathbf{H}_{\rm hf} ~|| ~[001]$ & $\mathbf{H}_{\rm hf}~ || ~[001]$ & no\\
  (4.4) & [1$\bar{1}$0] & [001] & dip & $\mathbf{H}_{\rm hf}~ ||~ [010]$ & $\mathbf{H}_{\rm hf}~ ||~ [100]$ & $\mathbf{H}_{\rm hf}~ ||~ [010]$  & \textbf{yes (e)}\\
\noalign{\smallskip}\hline
\end{tabular}
\end{center}
\end{table}

\subsection{Magnetic structure}

The magnetic structure is given by $\mathbf{S} = \mathbf{S_0}\cos[(\mathbf{Q}_0+\mathbf{Q}_i)\cdot\mathbf{r}]$,
where the antiferromagnetic wavevector $\mathbf{Q}_0= (\frac{\pi}{a}, \frac{\pi}{a}, \frac{\pi}{c})$ is commensurate with the lattice.
Neutron diffraction reports an incommensurate wavevector
$\mathbf{Q}_i= \frac{\pi}{a}(\delta,\delta,0)$ with spatial modulation $\sqrt{2}a/\delta \approx 12 a \approx 5.4$ nm in the $ab$ plane,
and ordered moment $\mathbf{S_0}$ at the Ce site.
This modulation is significantly shorter than the average inter-vortex distance of $\sim 14$ nm in a field of 10~T.\cite{Koutroulakis08}
So it does not support a picture of overlapping extended states from the vortex cores leading to magnetic ordering.
Similarly this specific $\mathbf{Q}_i$ would be inconsistent with a field along [100], if it was related to the alignment of vortices along the
[100] direction.
In our re-analysis of the NMR spectra, we consider four cases:
(1) $\mathbf{Q}_i ~||~ [100]$,
(2) $\mathbf{Q}_i ~||~ [010]$, and
(3) $\mathbf{Q}_i ~||~ [110]$,
(4) $\mathbf{Q}_i ~||~ [1\bar{1}0]$.
Case (1) was proposed by
Young et al.\cite{CurroCeCoIn5FFLO} for NMR measurements under the condition $\mathbf{H}_0 ~||~ [100]$.
Case (2) should be equally likely as case (1) because of the tetragonal symmetry of the crystal structure.
Case (3) was  proposed by Kenzelmann et al.\cite{KenzelmannCeCoIn5Qphase} for neutron diffraction measurements under the condition $\mathbf{H}_0~ ||~ [1\bar{1}0]$, and case (4) should be equally likely as case (3).
We have calculated the hyperfine fields for each of these cases for both purely isotropic and purely dipolar couplings, for $\mathbf{H}_0 ~ || ~ [100]$ with moments along both [100] and [001], and the results are summarized in Table (1).

\begin{figure*}
\begin{minipage}[t]{0.48\textwidth}
\centerline{  \includegraphics[width=1.05\textwidth]{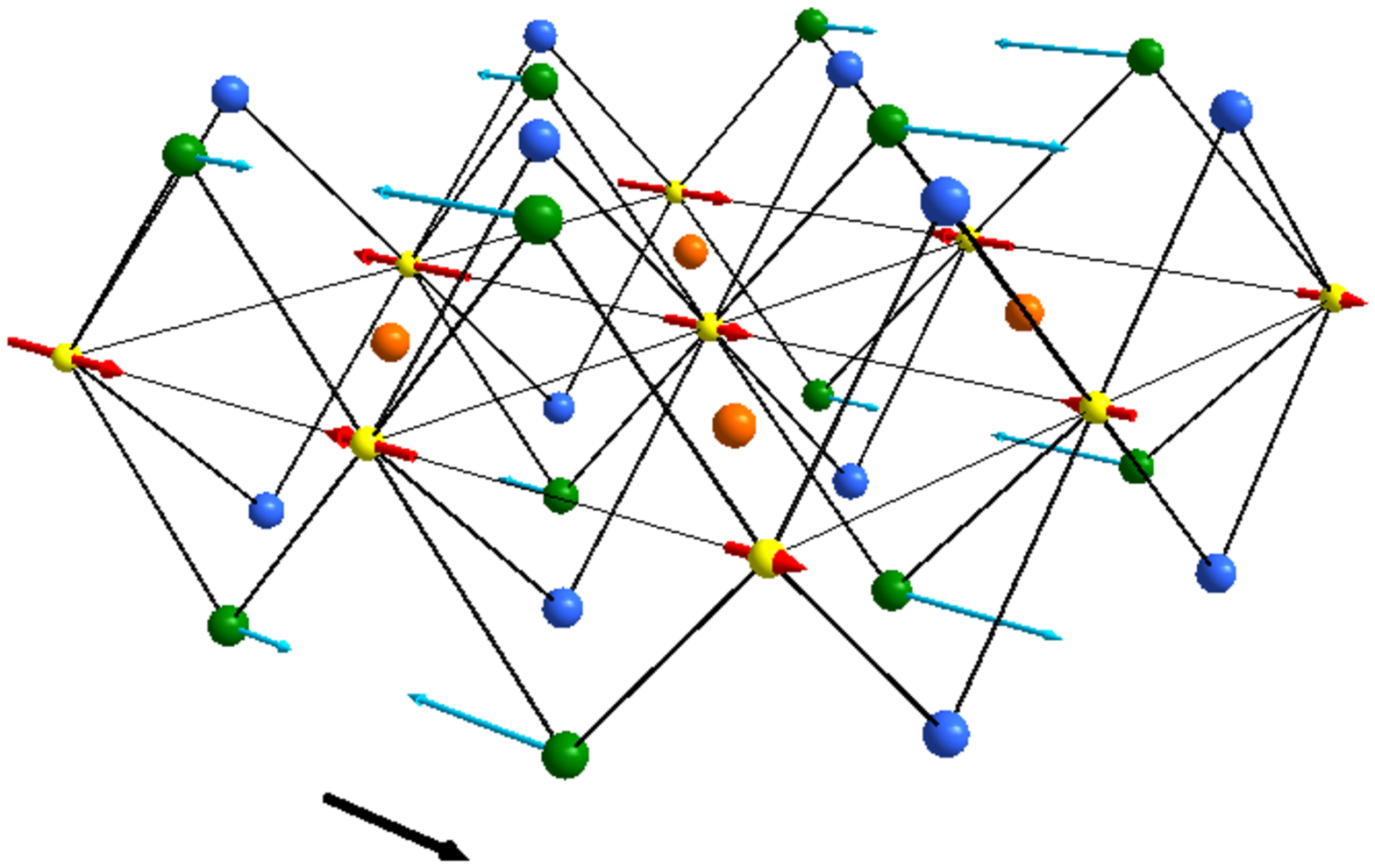} }
\caption{(Color online) Magnetic structure and hyperfine fields for isotropic hyperfine couplings to the In(2)
(\textbf{case (1.1)}).
The in-plane tetragonal structure is outlined in gray.
The Ce atoms are yellow, and their  moments are indicated by red arrows pointing along [100].
The In(1) atoms are orange, and the Co are not shown.
The In(2a) are  green and the In(2b) are blue.
The hyperfine fields at the In(2a) sites are indicated by blue arrows.
Here the hyperfine fields vanish at the In(1), Co and the In(2b) sites;
the direction of $\mathbf{H}_0$ is shown by the black arrow. }
\label{fig:isotropic}
\end{minipage}
\hfill
\begin{minipage}[t]{0.48\textwidth}
\begin{center}
\centerline{  \includegraphics[width=1.0\textwidth]{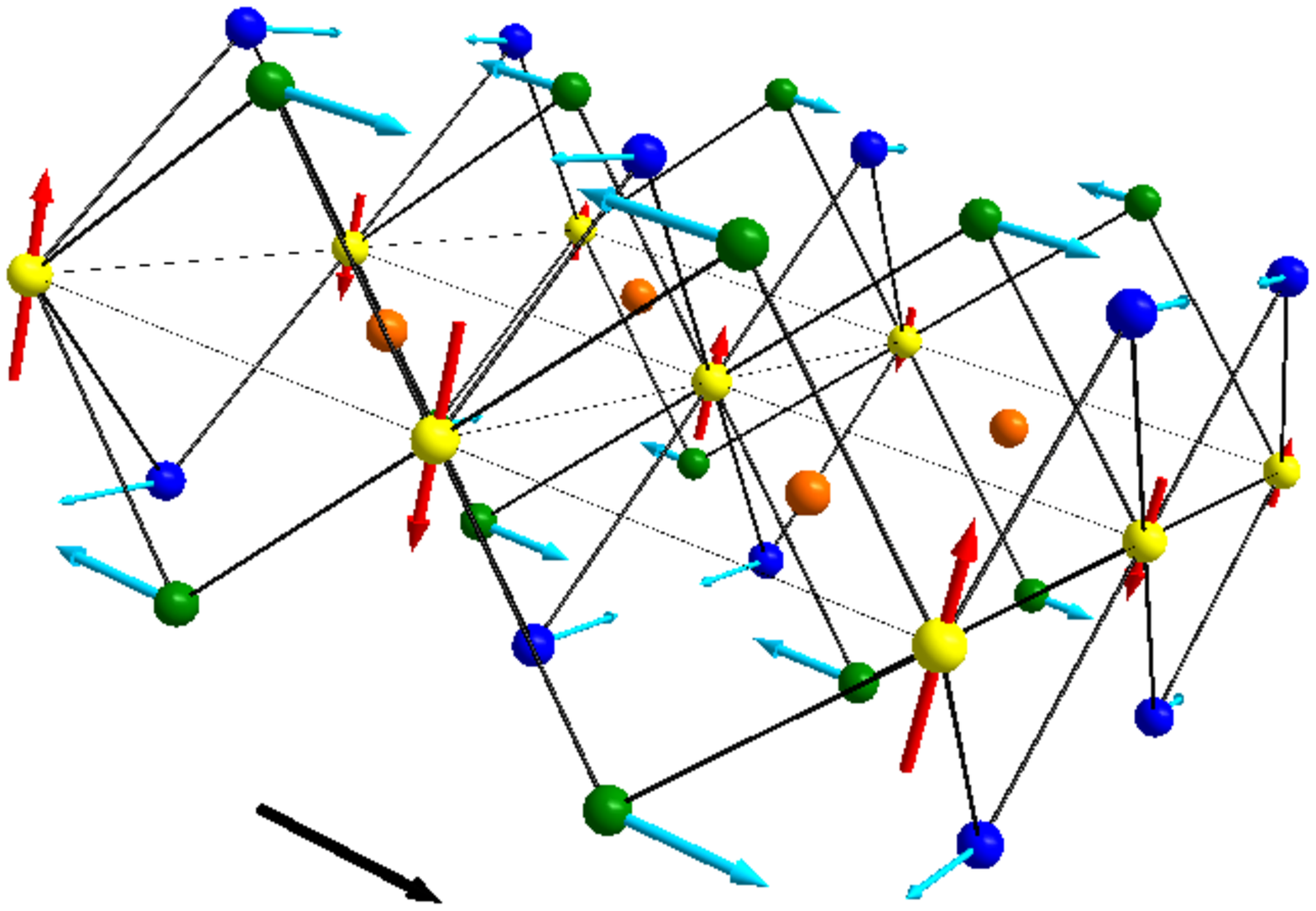} }
\end{center}
\caption{(Color online) Magnetic structure and hyperfine fields for dipolar hyperfine couplings to the In(2)
(\textbf{case (2.4)}). Same notation as in Fig.~\ref{fig:isotropic}.
The Ce moments point along [001].
Here the hyperfine fields vanish at the In(1) and Co sites, but not at the In(2) sites;
the direction of $\mathbf{H}_0$ is shown by the black arrow. }
\label{fig:dipolar}
\end{minipage}
\end{figure*}

  The cases that are most consistent with the NMR observations are (1.1) $\mathbf{Q}_i ~||~ [100]$, $\mathbf{S}_0 ~||~ [100]$ and isotropic coupling, (1.4) $\mathbf{Q}_i ~||~ [100]$, $\mathbf{S}_0 ~||~ [001]$ and dipolar coupling, and (2.4) $\mathbf{Q}_i ~||~ [010]$, $\mathbf{S}_0 || [001]$ and dipolar coupling. Cases (3.4) and (4.4) are consistent with the In(2) spectra, but give rise to an internal field at the In(1)
site that at first sight is inconsistent with experiment.
We will discuss this in more detail below in the discussion section.
Figures \ref{fig:isotropic} and \ref{fig:dipolar} show the magnetic structure and hyperfine fields for cases (1.1) and (2.4).
Case (1.1) is identical to the one originally proposed by Young and coworkers,\cite{CurroCeCoIn5FFLO}
which most likely will not minimize the magnetic contribution to the free energy, as the moments are either parallel or antiparallel to the applied field.
Cases (1.4), (2.4), (3.4) and (4.4), in which the moments are perpendicular to the applied field,
are physically more reasonable for antiferromagnetic ordering and agree with the neutron diffraction results of
$\mathbf{Q}_i \perp \mathbf{S}_0$ and $\mathbf{H}_0 \perp \mathbf{S}_0$.\cite{KenzelmannCeCoIn5Qphase}

\section{Discussion}

\subsection{Hyperfine constants from the Knight shift}

In \cecoin\ measurements of the Knight shift in the normal state show that the hyperfine couplings for the In(2a) site ($\mathbf{r}_k=( \pm \frac{a}{2},0,z_0 )$) for the applied field along (100), (010) and (001) are 10.3 kOe/$\mu_B$, 0.0 kOe/$\mu_B$, and 32.4 kOe/$\mu_B$, respectively.\cite{CurroAnomalousShift}
For this site, Equations (\ref{eqn:isotropic}) and (\ref{eqn:dipolar}) yield:
\begin{eqnarray}
K_a&=& (2B_{iso} + B_{dip}(1-3\cos2\theta_z))\chi_a\\
K_b&=& (2B_{iso} - 2B_{dip})\chi_b\\
K_c&=& (2B_{iso} + B_{dip}(1+3\cos2\theta_z))\chi_c,
\end{eqnarray}
where $\chi_{\alpha}$ is the susceptibility in the $\alpha$ direction.
Using the experimental numbers,\cite{CurroAnomalousShift}
we find $B_{iso}=B_{dip}=7.1$ kOe/$\mu_B$ and $\theta_z = 29^{\circ}$.  The difference between this angle and that of the
crystal structure ($\theta_z = 45^{\circ}$) probably is related to  details of the bonding of the In $4p$ orbitals, and will need further investigations.

\subsection{Hyperfine fields in the B phase}

\begin{figure*}
\begin{minipage}[t]{0.49\textwidth}
\centerline{  \includegraphics[width=\textwidth]{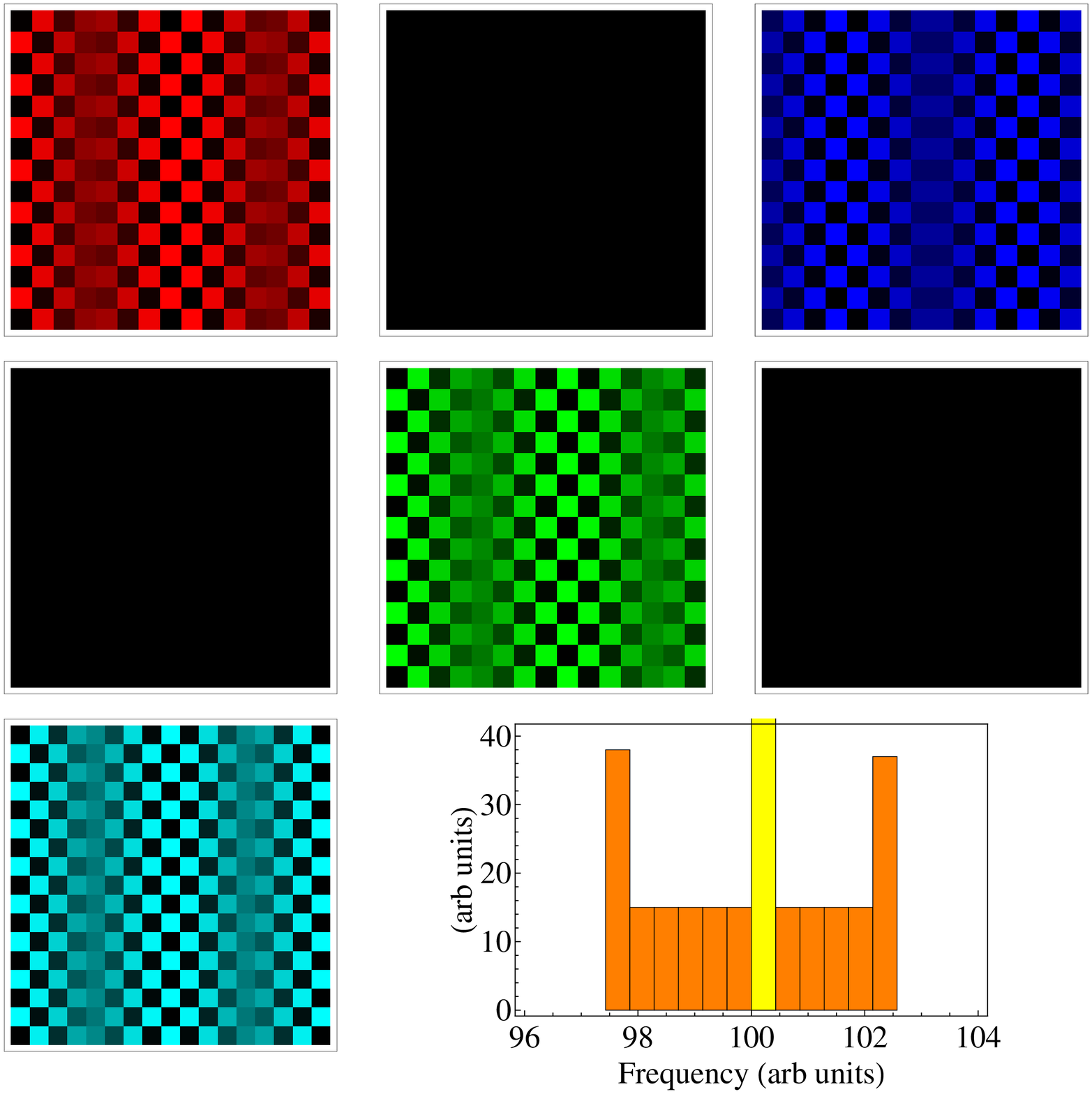} }
\caption{(Color online)
Real-space map (15 x 15 unit cells) of the hyperfine field at the In(2a) (Eq. 6) (upper row) and the In(2b) (Eq. 7) (middle row) in the $ab$ plane.
Shown along the horizontal are the components of the hyperfine field along the ($\hat{\mathbf{x}}$, $\hat{\mathbf{y}}$, $\hat{\mathbf{z}}$) directions in (red, green blue) shading,  for \textbf{case (2.4)} with $\mathbf{Q}_0 ~ || ~ [010]$ and $\mathbf{S}_0 ~||~[001]$. Black corresponds to zero hyperfine field.
The lower row shows the spin density along the [001] direction (cyan), and the histogram of resonant frequencies for the In(2a) (orange) and the In(2b) (yellow).}
\label{fig:hyperfinemap2p4}
\end{minipage}
\hfill
\begin{minipage}[t]{0.49\textwidth}
\centerline{  \includegraphics[width=\textwidth]{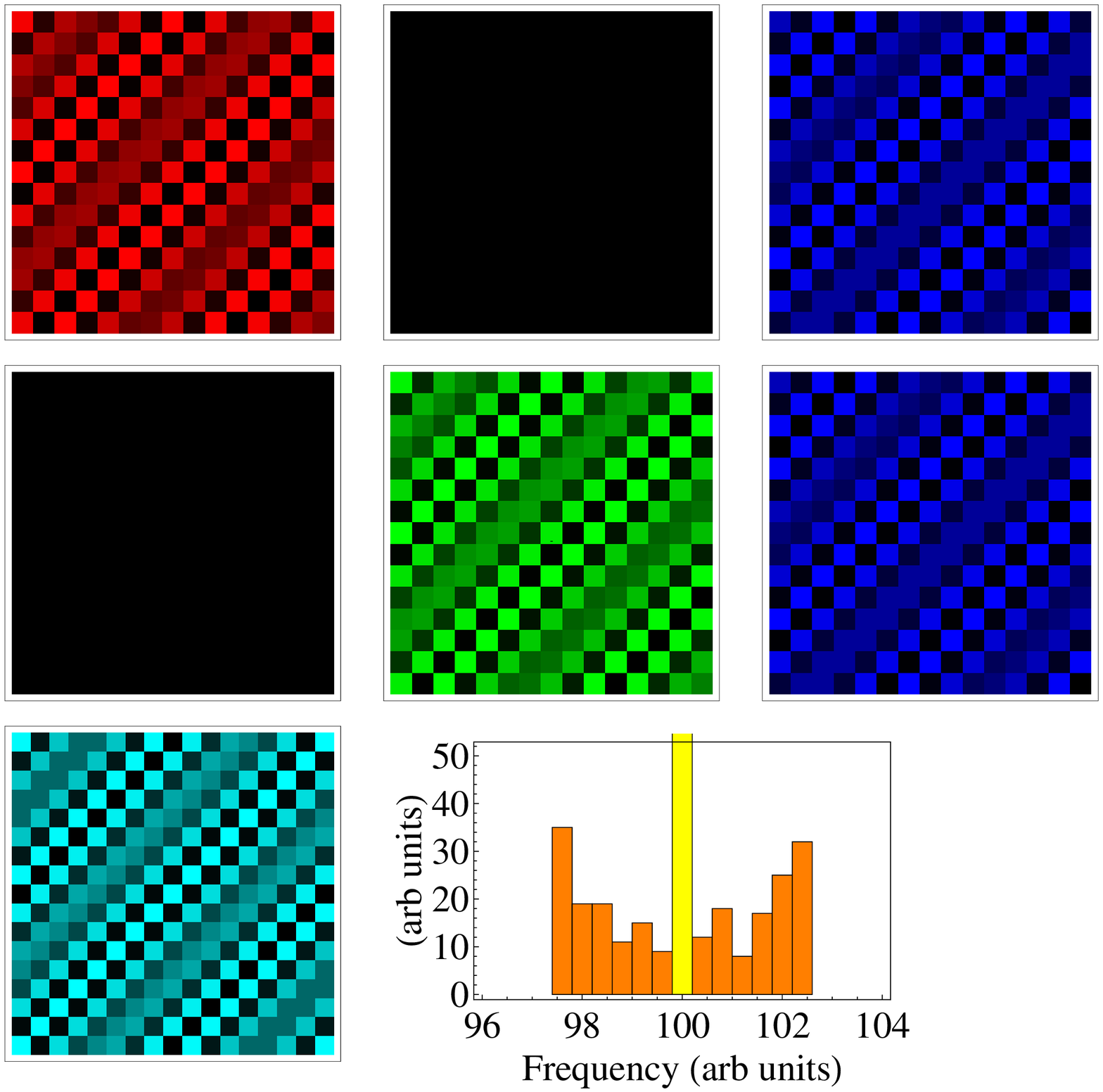} }
\caption{(Color online) The same notation as in Fig.~\ref{fig:hyperfinemap2p4},
but for \textbf{case (3.4)} with $\mathbf{Q}_0 ~ || ~ [110]$ and $\mathbf{S}_0 ~||~ [001]$.
The corresponding histogram of resonant frequencies is in better agreement with experiment.}
\label{fig:hyperfinemap3p4}
\end{minipage}
\end{figure*}

For cases (1.4), (2.4), (3.4) and (4.4) the hyperfine field at the In(2a) and In(2b) sites varies spatially
$\sim\cos[(\mathbf{Q}_0+\mathbf{Q}_i)\cdot\mathbf{r}]$ with modulus:
\begin{eqnarray}
\label{eqn:hypfield}
\mathbf{H}_{hf}(2a) &=& S_0B_{dip}[3\sin(2\theta_z)\cos(\frac{\pi\delta_x}{2})\hat{\mathbf{a}} + (1-\cos(2\theta_z)\sin(\frac{\pi\delta_x}{2})\hat{\mathbf{c}}]\\
\mathbf{H}_{hf}(2b) &=& S_0B_{dip}[3\sin(2\theta_z)\cos(\frac{\pi\delta_y}{2})\hat{\mathbf{b}} + (1-\cos(2\theta_z)\sin(\frac{\pi\delta_y}{2})\hat{\mathbf{c}}]
\end{eqnarray}
In each case, the hyperfine field at the In(2a) site oscillates along the modulation direction with a component along [100] and the resulting spectrum is described by Figs. \ref{fig:spectra}, \ref{fig:hyperfinemap2p4} and \ref{fig:hyperfinemap3p4}.
Using the values $\delta_x=\delta_y\approx 0.12$ and $S_0\approx 0.15\mu_B$ as reported by Kenzelmann et al.,\cite{KenzelmannCeCoIn5Qphase}
we find $h_{hf}^0 \approx 2.6$ kOe, which is about twice the experimental value of 1.3 kOe.
The difference may be related to uncertainties in the hyperfine coupling itself \cite{CurroKSA} or changes in the magnetic structure for the field along [100]. Note that the modulation and ordered moment may differ for field oriented along [100], in which case Eq.~(\ref{eqn:hypfield}) will give a different value. In fact the NMR result is consistent with the locus of points given by $\delta_x = \delta_{x0}\cos^{-1}(S_0^0/S_0-1)$, where $\delta_{x0} = 0.3$ and $S_0^0 = 0.14\mu_B$.

For cases (3.4) and (4.4), where the modulation is along [110] or [1$\bar{1}$0], the hyperfine field at the In(1) site does not cancel but has a component along the [001] direction.  This field can give rise to a minor shift and/or broadening of the In(1) line.  The spectra (Fig. \ref{fig:spectra}) clearly show that the In(1) line shifts and is only slightly broadened. However, the shift and broadening may come from the onset of spin shift suppression in the superconducting state and the presence of superconducting vortices.  Therefore, we cannot distinguish the presence of a hyperfine field from the antiferromagnetic structure at the In(1) site within experimental error.  As seen in Figs. \ref{fig:hyperfinemap2p4} and \ref{fig:hyperfinemap3p4}, the calculated spectra for case (3.4) is closer to the experimental one.
We speculate that the true magnetic structure for the field $\mathbf{H}_0$ along [100] is best described either by case (2.4),
(3.4) or (4.4) with  $\mathbf{H}_0 \perp \mathbf{S}_0$, which will minimize the
free energy of the antiferromagnet.
In each case the hyperfine field at the In(2a) and In(2b) have components perpendicular to $\mathbf{H}_0$,
but these components only give rise to small shifts of the resonant frequency that are difficult to distinguish from the Knight shift.  The crucial element is that the hyperfine field at the In(2a) is along $\mathbf{H}_0$.
\vspace{-0.2cm}

\subsection{Nature of the B phase}

The fact that the antiferromagnetism exists only in field and only within the superconducting phase indicates a strong coupling between these order parameters.
Kenzelmann and coauthors\cite{KenzelmannCeCoIn5Qphase} analyzed the symmetry of the superconducting state for such a coupling and concluded that the superconducting order parameter,
$\Delta_{\mathbf{Q}}$ acquires the finite momentum $\mathbf{Q}$ of the antiferromagnetic order parameter, ${\mathbf{M_Q}}$.
This corresponds to a modulation of the order parameter in real space that presumably is out-of-phase with the antiferromagnetism.
In other words, the antiferromagnetic order is maximum at the nodes of $\Delta_{\mathbf{Q}}$.
Since then various microscopic models have been proposed to explain the field induced antiferromagnetic order.\cite{Yanase09, Kee09}
Curiously, this scenario is similar to that observed in the ferropnictide SrFe$_2$As$_2$ under pressure.\cite{takigawaSrFe2As2pressure}
In this compound, a novel hybrid state of coexisting superconductivity and antiferromagnetism emerges above 5 GPa.  We speculate that these two novel states may in fact be the same.  Although superconductivity and antiferromagnetism are known to coexist inhomogeneously in a number of \textit{doped} high T$_c$, heavy fermion, and ferropnictide systems, the highly clean undoped CeCoIn$_5$ and SrFe$_2$As$_2$ materials support the emergence of this fragile but \textit{intrinsic} thermodynamic phase of modulated antiferromagnetism and superconductivity.  In CeCoIn$_5$, this phase is quickly destroyed by doping and is replaced by a commensurate order at zero field for sufficiently high Cd doping.\cite{YoshiCdFFLO,CeCoIn5CdDroplets,ricardomultipleT1,nicklasPHTdiagramCeCoIn5}
Clearly many questions about this new state of matter remain unexplained, such as the driving mechanism(s), the origin of the incommensurate wavevector, and the nature of the excitations.

\section{Conclusions}

In summary, we have shown that both NMR and neutron diffraction measurements in the field-induced B phase of \cecoin\ are consistent
with magnetic structures where
$\mathbf{Q}_i \perp \mathbf{S}_0$ and $\mathbf{H}_0 \perp \mathbf{S}_0$.
The incommensurate modulation $\textbf{Q}_i$ lies possibly along either [010] or [110] direction for magnetic field pointing along [100] in real space.
Tetragonal equivalent directions for $\textbf{Q}_i$, [100] and [1$\bar{1}$0], are also possible.
Based on our analysis of the NMR spectra, we speculate that the B phase of \cecoin\ represents an intrinsic phase of modulated superconductivity and antiferromagnetism
that can only emerge in a highly clean system. Further NMR and neutron diffraction measurements are necessary for the same field orientations to unravel the origin
of the field-induced antiferromagnetic structure.

\begin{acknowledgements}
We would like to thank R.\ Movshovich, V.\ Mitrovi\'c, and M.\ Kenzelmann for valuable discussions and sharing their results.
Work at Los Alamos National Laboratory was performed under the
auspices of the US Department of Energy under grant no.\ DE-AC52-06NA25396.
\end{acknowledgements}




\end{document}